\documentclass[psfig,aps,twocolumn]{revtex4}
\usepackage{amssymb}
\usepackage{amsmath}
\usepackage{graphicx}
\usepackage{lscape}
\usepackage{booktabs}
\begin{document}
\title{Towards a qualitative hadron-parton correspondence \\ in the
       nuclear modification factor}
\author{Ben-Hao Sa$^{1,2}$ \footnote{sabh@ciae.ac.cn},
        Dai-Mei Zhou$^2$\footnote{zhoudm@mail.ccnu.edu.cn},
        Yu-Liang Yan$^{1,2}$ \footnote{yanyl@ciae.ac.cn},
        Wen-Di Liu$^1$, Shou-Yang Hu$^1$, Xiao-Mei Li$^1$,
	Liang Zheng$^3$, Gang Chen$^3$, and Xu Cai$^2$}
\affiliation{$^1$ China Institute of Atomic Energy, P. O. Box 275 (10),
                  Beijing, 102413 China. \\
             $^2$ Key Laboratory of Quark and Lepton Physics (MOE) and Institute
                  of Particle Physics, Central China Normal University, Wuhan
                  430079, China. \\
             $^3$ Physics Department, China University of Geoscience, Wuhan,
                  430074, China.}
\begin{abstract}
In this work, we propose a method to show the correspondence between hadron
and its quark component nuclear modification factors. A parton and hadron
cascade model PACIAE based on the PYTHIA6.4 is employed to calculate the
hadron and its quark component nuclear modification factors in the 0-5\%
most central $Pb+Pb$ collisions at $\sqrt{s_{NN}}$=2.76 TeV. It turns out
that the hadron nuclear modification factor is usually smaller than that
of its quark component. On the other hand, it is shown in our study that
the ``dead cone effect" is more likely to be identified with the quarks and
mesons but not with the baryon states obviously.
\end{abstract}
\maketitle

\section {Introduction}
Jet quenching (energy loss) and particle azimuthal asymmetries (elliptic flow
etc.) are the essential probes investigating the quark-gluon plasma (QGP) in
the ultra-relativistic nucleus-nucleus collisions at RHIC
\cite{rhic1,rhic2,rhic3,rhic4} and LHC \cite{ALICE1,CMS1,ATLAS1} energies.
Nuclear modification factor is an important measurement exploring the jet
quenching effect \cite{shen,wang}. The parton nuclear modification factor is
responsible for the energy loss in partonic stages of initial state and parton
rescattering stage. And the hadron nuclear modification factor in final
hadronic state is one measure of the jet quenching in the hadronic stages of
hadronization and hadronic rescattering. It is speculated that a connection
can be built with the measurable final state hadron nuclear modification factor
and the underlying quark energy loss.

The hadron transverse momentum dependent nuclear modification factor in
final hadronic state of the nucleus-nucleus ($A+A$) collision is usually
defined as \cite{cms2}
\begin{equation}
R_{AA}^{h}(p_T)=\frac{[dY(h)/dp_T]_{AA}}
                          {N_{bin}[dY(h)/dp_T]_{pp}},
\label{equ1}
\end{equation}
where $Y(h)$ stands for the hadron ($h$) yield and $N_{bin}$ refers to the
binary collision number usually obtained within the optical Glauber model
and/or Monte Carlo Glauber model \cite{shor,abel,abel1,misk,mill,phobos,loiz}.
It is sensitive to the relative variation of the particle transverse momentum
distribution in the nucleus-nucleus and proton-proton collisions. However,
the transverse momentum dependent nuclear modification factor is dominated by
the energy loss (jet quenching). A substantial suppression of $R_{AA}^h(p_T)
<1$ in the intermediate to high $p_T$ region is anticipated as the signal of
the partonic energy loss in Quark Gluon Matter (QGM) \cite{wilk}. The behavior
of nuclear modification factor as a function of $p_T$ is not monotonous. In
general, a peak appears at $p_T\sim$ 2 GeV/c, which the valley follows at the
$p_T\sim$ 6 GeV/c, then the function increases monotonously toward the unity
\cite{cms2}. Thus the model simulation is hard to reproduce the
$R_{AA}^h(p_T)$ calculated with experimentally measured transverse momentum
distributions in $p-p$ and nucleus-nucleus collisions, over full $p_T$ region
especially.

Similarly, the quark transverse momentum dependent nuclear modification factor
in the deconfined QGM reads
\begin{equation}
R_{AA}^{q}(p_T)=\frac{[dY(q)/dp_T]_{AA}}
                          {N_{bin}[dY(q)/dp_T]_{pp}}.
\label{equ2}
\end{equation}

Taken $\Lambda^0(uds)$ hadron as an example, it is consisted of the constituent
quarks $u$, $d$, and $s$. Identifying the correspondence between the hadron
and its quark component nuclear modification factors would be worthwhile to
study the effects of jet quenching. The recombination (coalescence) model
\cite{yang,ko,bass} is hard in constructing this correspondence due to the
complication in dealing with the flavor composition of constituent quarks. The
spirit of ``correspondence principle" \cite{dict,wold,lim} has been inspiring
us to construct the correspondence between hadron and its quark component
nuclear modification factors based on physical deductions. Consequently, a
parton and hadron cascade model of PACIAE is employed calculating the hadron
nuclear modification factor in final hadronic state (FHS) and its quark
component nuclear modification factor in partonic state after parton-parton
rescattering (PSw/R).

\section {Physical deductions}
The hadron ($h$) normalized transverse momentum distribution
\begin{equation*}
\frac{1}{N(h)}dY(h)/dp_T
\end{equation*}
corresponds to its quark component normalized transverse momentum
distribution as
\begin{equation*}
\frac{1}{N_{cq}}[\sum\limits_q \frac{1}{N(q)}dY(q)/dp_T].
\end{equation*}
In the above expressions the $N(h)$ ($N(q)$) is total multiplicity of the
hadron $h$ (quark $q$). $N_{cq}$ denotes the number of constituent quarks,
and the sum is taken over all the constituent quarks. If both are multiplied
by $N(h)$, one can get the resulting hadron ($h$) un-normalized transverse
momentum distribution
\begin{equation*}
dY(h)/dp_T,
\end{equation*}
which appears in the Eq. (\ref{equ1}) and is corresponding to
\begin{equation*}
\frac{1}{N_{cq}}[\sum\limits_q \frac{N(h)}{N(q)}dY(q)/dp_T].
\end{equation*}
Then we have the hadron nuclear modification factor of Eq. (\ref{equ1})
corresponding to its quark component nuclear modification factor of
\begin{equation}
R_{AA}^{h(qc)}(p_T)=\frac{[\sum\limits_q {w_qdY(q)/dp_T}]_{AA}}
                        {N_{bin}[\sum\limits_q {w_qdY(q)/dp_T}]_{pp}},
\label{equ3}
\end{equation}
where the superscript $h(qc)$ denotes the quark component of hadron $h$ and
$w_q=\frac{N(h)}{N(q)}$.

The quark energy loss is supposed to follow the so called ``dead cone effect"
\cite{malt} and the size of dead cone is proportional to the mass of the QED
and/or QCD emitter. Quark with smaller mass will lose more energy during the
propagation through the dense medium. Therefore, one may expect the flavor
ordering of the energy loss and of nuclear modification at the parton level\cite{wilk,djor}:
\begin{equation*}
\Delta E(g)>\Delta E(u)>\Delta E(s)>\Delta E(c)>\Delta E(b)> \Delta E(t)
\end{equation*}
and
\begin{equation}
R_{AA}^g<R_{AA}^u<R_{AA}^s<R_{AA}^c<R_{AA}^b<R_{AA}^t.
\end{equation}
We also investigate the ``dead cone effect" in the partonic state after
parton-parton rescattering and in final hadronic state in the 0-5\%
centrality $Pb+Pb$ collisions at $\sqrt{s_{NN}}$=2.76 TeV in this paper.

\begin{widetext}
\begin{center}
\begin{figure}[htbp]
\centering
\hspace{-2.10cm}
\includegraphics[width=0.45\textwidth,angle=270]{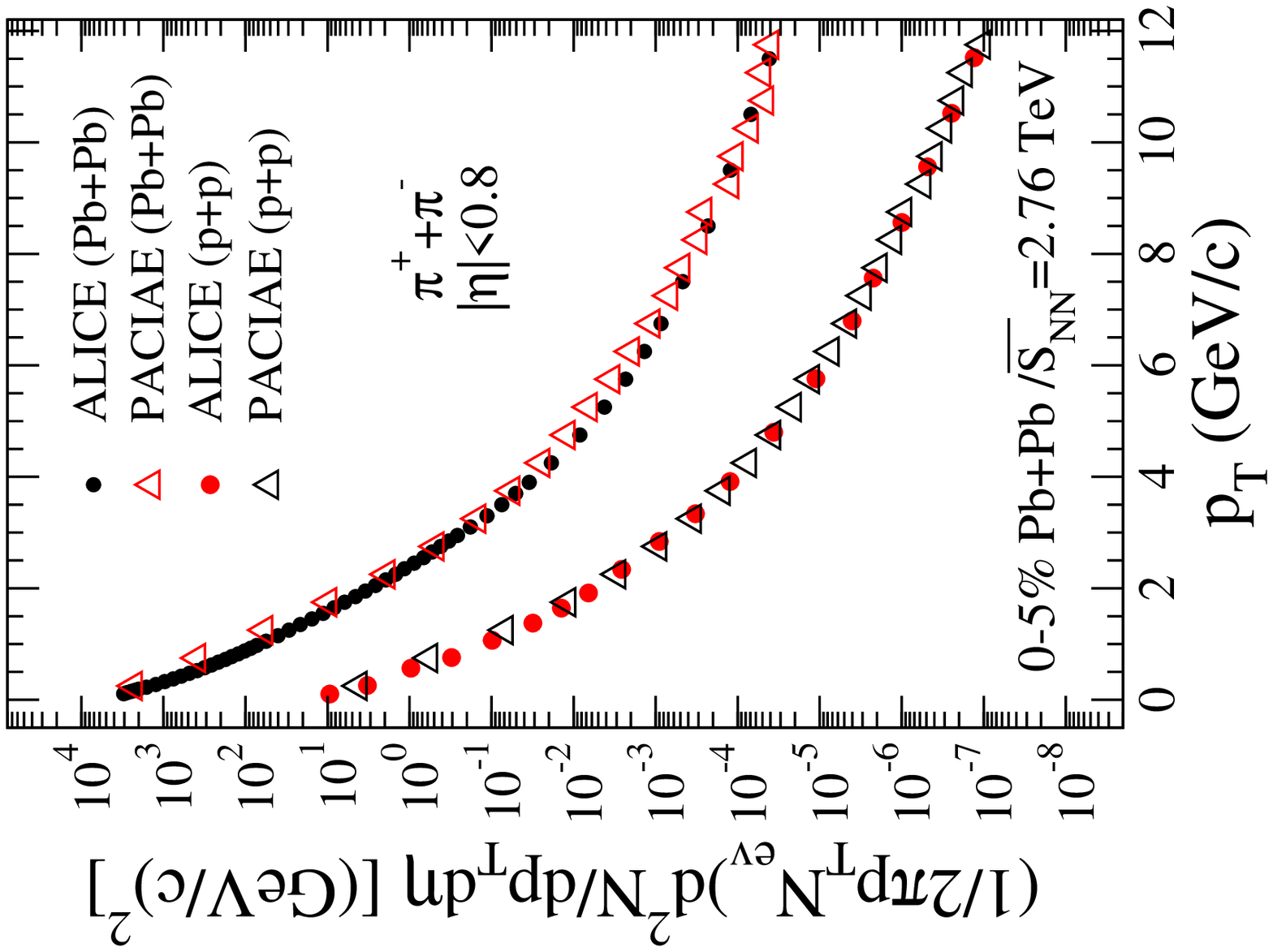} \hspace{-2.1cm}
\includegraphics[width=0.45\textwidth,angle=270]{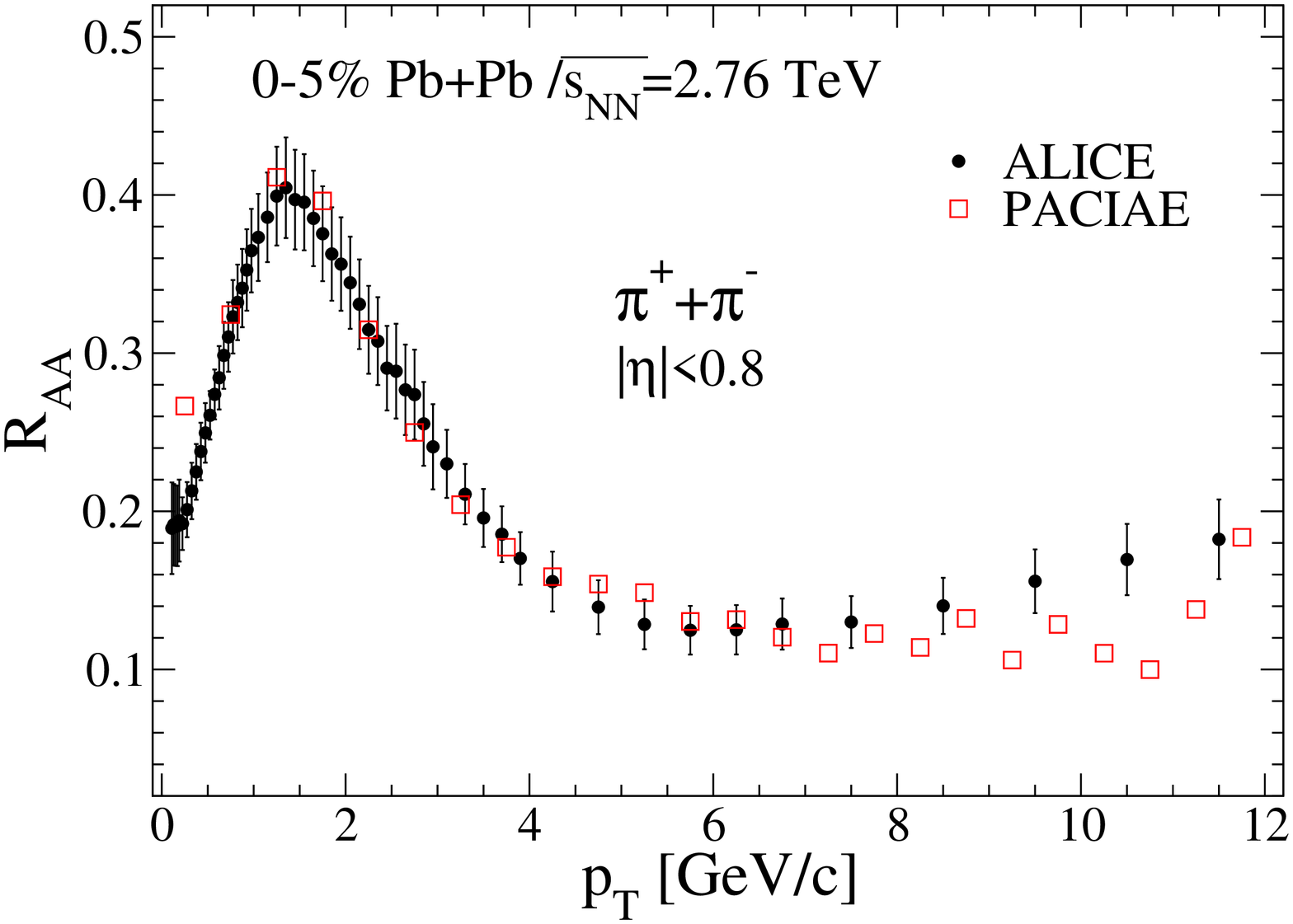}
\caption{The left panel shows $\pi^++\pi^-$ transverse momentum distributions
in FHS of 0-5\% most central $Pb+Pb$ and $p+p$ collisions at $\sqrt{s_{NN}}$=
2.76 TeV (ALICE data are taken from \cite{alice3}). The right panel is
$R_{AA}^{\pi^++\pi^-}(p_T)$.}
\label{pt_raa}
\end{figure}
\end{center}
\end{widetext}

\begin{widetext}
\begin{center}
\begin{figure}[htbp]
\centering
\includegraphics[width=0.90\textwidth,angle=270]{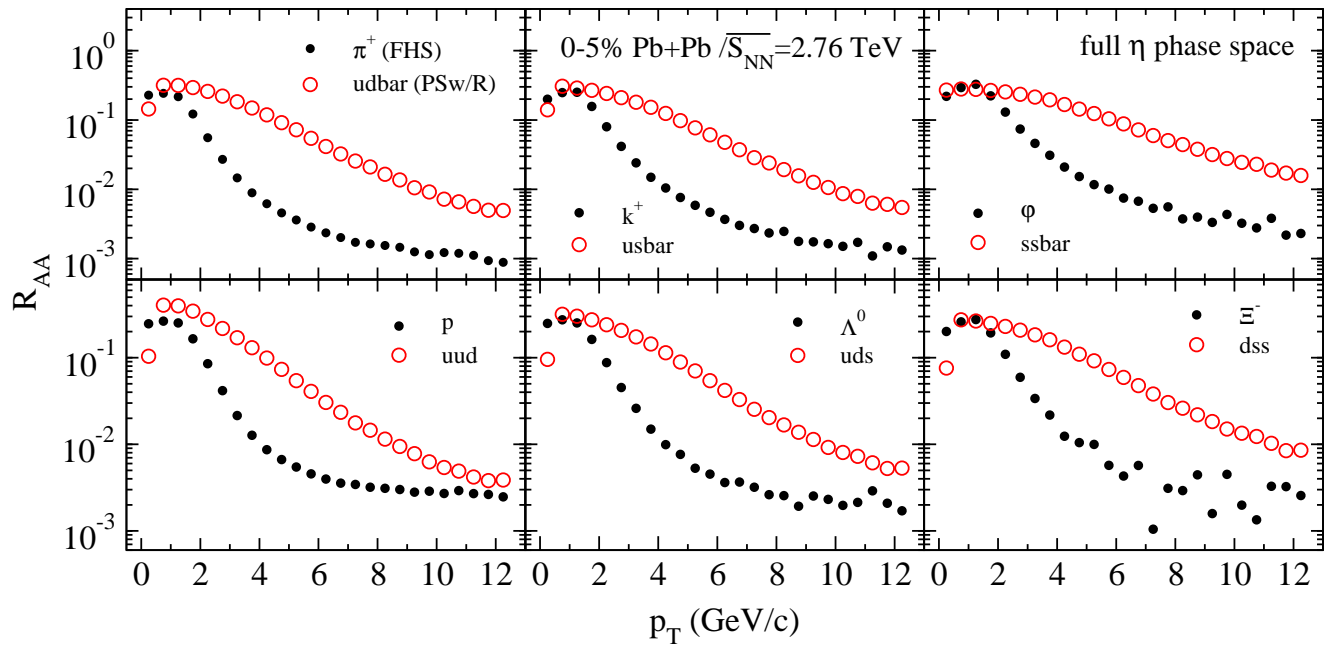}
\caption{The nuclear modification factor of mesons and baryons (in FHS) as well
as their quark component ones (in PSw/R) in 0-5\% most central $Pb+Pb$ at
$\sqrt{s_{NN}}$=2.76 TeV.}
\label{raa2_m}
\end{figure}
\end{center}
\end{widetext}

\section {PACIAE model}
The parton and hadron cascade model PACIAE \cite{sa1} based on PYTHIA6.4
\cite{sjos1}, is employed to calculate the hadron and its quark component
nuclear modification factors in the 0-5\% most central $Pb+Pb$ collisions at
$\sqrt{s_{NN}}$=2.76 TeV. In PACIAE model the nucleon initial position in
a nucleus-nucleus collision is distributed randomly according to the
Woods-Saxon distribution and the number of participant (spectator) nucleons
determined by the Glauber model \cite{shor,abel,abel1,misk,mill,phobos,loiz}.
Together with the initial momentum setup of $p_x=p_y=0$ and $p_z=p_{beam}$,
the initial nucleon list is constructed for a colliding nucleus-nucleus
system. Collision happens between a pair of two nucleons if their relative
transverse distance is less than or equal to the minimum approaching distance:
$D\leq\sqrt{\sigma_{NN}^{tot}/\pi}$. The collision time is calculated with the
assumption of straight-line trajectories. All such nucleon pairs compose the
nucleon-nucleon ($NN$) collision (time) list. A $NN$ collision with least
collision time is selected from the collision list and executed by PYTHIA6.4
(subroutine PYEVNW) with the string hadronization temporarily turned-off and
the strings as well as diquarks broken-up. The nucleon list and $NN$ collision
list are then updated. A new $NN$ collision with least collision time is
selected from the updated $NN$ collision list and executed by PYTHIA6.4.
Such a routine is repeated until the $NN$ collision list is empty. The initial
partonic state (IPS) of a nucleus-nucleus collision is reached.

\begin{figure}[htbp]
\centering
\includegraphics[width=0.45\textwidth,angle=270]{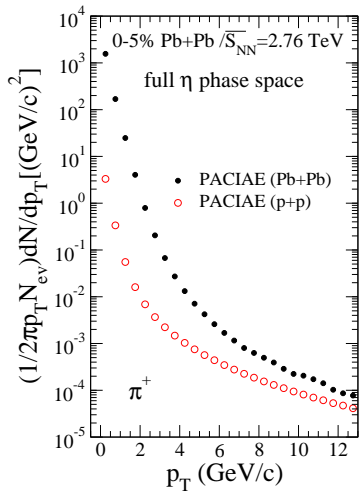}
\caption{$\pi^+$ transverse momentum distribution in full pseudorapidity phase
space in FHS of the 0-5\% most central $Pb+Pb$ collisions at $\sqrt{s_{NN}}$
=2.76 TeV.}
\label{pt_full}
\end{figure}

The above initial partonic state is then proceeding to a partonic rescattering
where the LO-pQCD parton-parton cross section \cite{ranft,field} are
employed. The state, after partonic rescattering is refereed to as partonic
state after rescattering (PSw/R).

After string recovering, the Lund string fragmentation regime is employed to
hadronize the strings resulting an intermediate hadronic state. This
intermediate hadronic state proceeds with the hadronic rescattering.
A final hadronic state (FHS) is reached for a nucleus-nucleus collision
eventually.

\section {Results and conclusions}
In the PYTHIA model a $K$ factor is introduced multiplying the hard scattering
cross section. Together with the Lund string fragmentation parameters of
$\alpha$ and $\beta$ as well as the Gaussian width ($\omega$) of primary
hadron transverse momentum distribution are adjusted fitting globally the
ALICE data \cite{alice3} of $\pi^++\pi^-$ $p_T$ distributions in the 0-5\%
most central $Pb+Pb$ and $p+p$ collisions at $\sqrt{s_{NN}}$=2.76 TeV. The
fitted parameters: $K=$2.7, $\alpha=$1.3, $\beta=$0.09, and $\omega=$0.575
for $Pb+Pb$ ($K=$0.7, $\alpha=$0.3, $\beta=$0.58, and $\omega$=0.36 for $p+p$)
are used in the later calculations, where the full $\eta$ phase space is
assumed. The hadron $p_T$ distribution in heavy-ion collisions is observed to
transit from exponential like at low transverse momenta to power law like at
the high transverse momenta \cite{alice3}. It is approximately assumed by
multiplying the exponentially simulated primary hadron $p_T$ with a factor
of 1.8 if $p_T$ is larger than 2.5 GeV/c in the PYTHIA model (subroutine
PYPTDI) for $Pb+Pb$ collisions.

\begin{widetext}
\begin{center}
\begin{figure}[htbp]
\centering
\includegraphics[width=0.90\textwidth,angle=270]{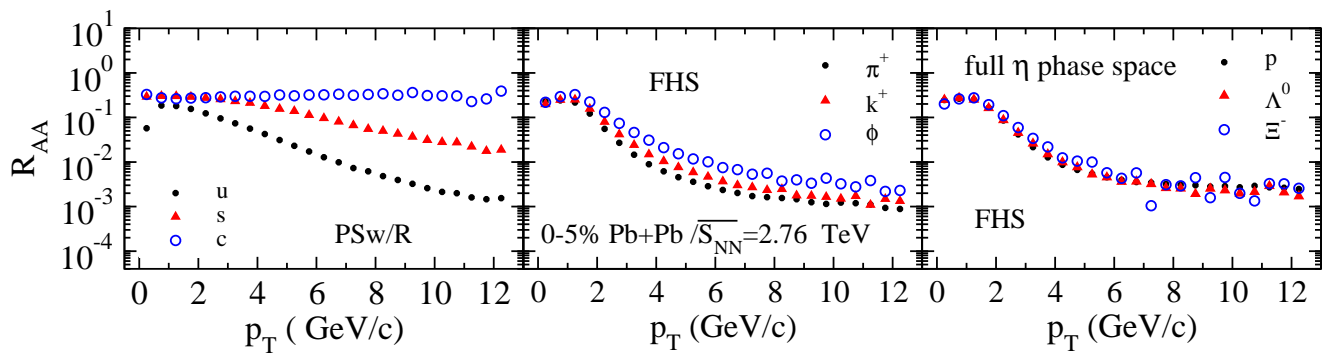}
\caption{The nuclear modification factor of quarks (left, in PSw/R), mesons
(middle, in FHS), and baryons (right, in FHS) in 0-5\% most central $Pb+Pb$
collisions at $\sqrt{s_{NN}}$=2.76 TeV.}
\label{raa2_qmb}
\end{figure}
\end{center}
\end{widetext}

We see in the left panel of Fig.(\ref{pt_raa}) that the fitted $\pi^++\pi^-$
$p_T$ distributions in the 0-5\% most central $Pb+Pb$ (black open
triangles-up) and $p+p$ (red open triangles-up) collisions at $\sqrt{s_{NN}}$
=2.76 TeV are well consistent with the ALICE data (black full circles for
$Pb+Pb$ and red full circles for $p+p$) \cite{alice3}, respectively. The
theoretical $R_{AA}^{\pi^++\pi^-}(p_T)$ (red open squares, calculated with the
fitted $p_T$ distributions) is also found to agree with the experimental one
(black full circles, calculated with the ALICE $p_T$ distribution data)
within error bars.

The calculated mesons $R_{AA}(p_T)$ in full pseudorapidity phase space in FHS
(black solid circles) are compared with their quark components ones in PSw/R
(red open circles) in the upper panels of Fig.(\ref{raa2_m}): left panel for
$\pi^+$, middle for $K^+$, and right for $\phi$, respectively. The lower
panels are the same as upper ones, but showing baryon sectors of $p$
(lower-left panel), $\Lambda^0$ (lower-middle), and $\Xi^-$ (lower-right),
respectively. One can observe in the upper (lower) panel that, in $p_T$ region
above $p_T\sim$2 GeV/c the meson (baryon) $R_{AA}$ is generally less than its
quark component one. It can be understood as the hadron objects suffer more
energy loss than their quark components. In other words, there is follow-up
relation between hadron and its quark component energy loss. This observation
offers a possibility to distinguish the partonic and hadronic energy losses
as well as a reference for the reliable evaluation of the hadron nuclear
modification factor.

A special feature seen in Fig.(\ref{raa2_m}) is that the hadron $R_{AA}$
calculated with single differential $p_T$ distribution in full pseudorapidity
phase space is not increasing toward unity at high $p_T$ region as usually
observed in finite pseudorapidity phase space (cf. ALICE in $|\eta|<0.8$
\cite{alice4} and CMS in $|\eta|<1$ \cite{CMS1}). To understand this feature
we draw $\pi^+$ single differential $p_T$ distribution in full pseudorapidity
phase space in the 0-5\% most central $Pb+Pb$ and $p+p$ collisions at
$\sqrt{s_{NN}}$=2.76 TeV in Fig.(\ref{pt_full}). One sees here that the high
$(P_T)$ suppression in $Pb+Pb$ collision is stronger with $p_T$ increasing,
which is responsible for the $\pi^+$ $R_{AA}$ behavior at high $p_T$ region
in the full pseudorapidity phase space as shown in upper-left panel of
Fig.(\ref{raa2_m}).

In the Fig.(\ref{raa2_qmb}) we compare the calculated nuclear modification
factor for quarks ($u$, $s$, and $c$ in PSw/R, left panel), mesons ($\pi^+$,
$K^+$, and $\phi$ in FHS, middle panel), and baryons ($p$, $\Lambda^0$, and
$\Xi^-$ in FHS, right panel) in the 0-5\% most central $Pb+Pb$ collisions at
$\sqrt{s_{NN}}$=2.76 TeV. This figure shows that in $p_T$ region above
$p_T\sim$2 GeV/c the ``dead cone effect" is globally held for the sector of
quark (in PSw/R) and meson (FHS), but not showing obviously for baryons (FHS)
which has to be studied further. The nearly flat $R_{AA}^c(p_T)$ (cf. left
panel) is because of the $c$ quark is produced in the hard scattering
processes. Thus the $c$ quark $p_T$ distribution in $Pb+Pb$ collision is
nearly parallel to the one in $p+p$ collision at the same energy.

In summary, we introduce a correspondence between hadron nuclear modification
factor and its quark component one based on the spirit of correspondence
principle. Both of the hadron and its quark component nuclear modification factors are
then calculated separately with the parton and hadron cascade model PACIAE
based on PYTHIA6.4 in the 0-5\% most central $Pb+Pb$ collisions at
$\sqrt{s_{NN}}$=2.76 TeV. Because of the follow-up relation the hadron suffers
more energy loss than its quark component, the hadron $R_{AA}$ is less than
its quark component one in the $p_T$ region above $p_T\sim$2 GeV/c. A
comparison between the two serves a reference for the reliable evaluation of
hadron nuclear modification factor and a possibility to distinguish the
partonic and hadronic energy losses. Additionally, it turns out that the mass
hierarchy of $R_{AA}$ out of ``dead cone effect", can be found for the section
of quarks and mesons but not obviously for baryons. This has to be studied
further.

We thank C. B. Yang for discussions. This work was supported by the National
Natural Science Foundation of China (11775094, 11905188, 11775313), the
Continuous Basic Scientific Research Project (No.WDJC-2019-16), National Key
Research and Development Project (2018YFE0104800) and by the 111 project of
the foreign expert bureau of China.

\end{document}